\documentclass[11pt]{article}
\input epsf.tex
\newdimen\SaveWidth \SaveWidth=\textwidth
\newdimen\SaveHeight \SaveHeight=\textheight
\advance\SaveWidth by -\textwidth
\advance\SaveHeight by -\textheight

\divide\SaveWidth by 2
\divide\SaveHeight by 2
\advance\hoffset by \SaveWidth
\advance\voffset by \SaveHeight

\def\ie{\it i.e.}
\def\eg{\it e.g.}
\def\edm{\it edm}
\def\etc{\it etc.}
\def\starprod{\star~\rm{product}}
\def\nc{\rm noncommutative}
\def\ncg{\rm noncommutative~geometry}

\def\cpviol{CP ~\rm{violation}}
\def\cpviolng{CP ~\rm {violating}}

\def\etal{{\it et~al.}}

\def\abs#1{\left| #1\right|}

\let\badcite=\cite
\def\cite{~\badcite}

\def\slashchar#1{\setbox0=\hbox{$#1$}           
   \dimen0=\wd0                                 
   \setbox1=\hbox{/} \dimen1=\wd1               
   \ifdim\dimen0>\dimen1                        
      \rlap{\hbox to \dimen0{\hfil/\hfil}}      
      #1                                        
   \else                                        
      \rlap{\hbox to \dimen1{\hfil$#1$\hfil}}   
      /                                         
   \fi} 
    \def\slashword#1{\setbox0=\hbox{$#1$}        
  \dimen0=\wd0                                   
   \setbox1=\hbox{/} \dimen1=\wd1                
   \ifdim\dimen0>\dimen1                         
      \rlap{\hbox to \dimen0{\hfil\bf---\hfil}} %
      #1                                         %
   \else                                         
      \rlap{\hbox to \dimen1{\hfil$#1$\hfil}}    
      /                                          
    \fi}                                         %

\catcode`@=11
\newdimen\vbigd@men                             

\def\vbig#1#2{{\vbigd@men=#2\divide\vbigd@men by 2%
   \hbox{$\left#1\vbox to \vbigd@men{}\right.\n@space$}}}
\catcode`@=12

\catcode`@=11
\def\citenum#1{\csname b@#1\endcsname}
\catcode`@=12

\def\dofig#1#2{\centerline{\epsfxsize=#1\epsfbox{#2}}}
\begin{document}
\begin{titlepage}
\rightline{LBNL-47750}

\bigskip\bigskip

\begin{center}{\Large\bf\boldmath
CP-violation from Noncommutative Geometry\footnotemark \\}
\end{center}
\footnotetext{ This work was supported by the Director, 
Office of Science, Office
of Basic Energy Services, of the U.S. Department of Energy under
Contract DE-AC03-76SF0098.
}
\bigskip
\centerline{\bf I. Hinchliffe and N. Kersting}
\centerline{{\it Lawrence Berkeley National Laboratory, Berkeley, CA}}
\bigskip

\begin{abstract}

	If the geometry of space-time is $\nc$, $\ie$ $[x_{\mu},x_{\nu}]=i \theta_{\mu \nu}$,
 	then $\nc$ $\cpviolng$ effects may be manifest at low energies. For a 
	$\nc$ scale $\Lambda \equiv \theta^{-1/2} \leq 2~~TeV$, $\cpviol$ from $\ncg$ 
	is comparable to that from the Standard Model (SM) alone: the $\nc$ 
	contributions to $\epsilon$ and $\epsilon'/\epsilon$ in the $K$-system,
	may actually dominate over
	the Standard Model contributions. Present data permit $\ncg$ to be the only source
	of $\cpviol$. Furthermore the most recent findings
	for $g-2$ of the muon are consistent with predictions from $\ncg$.

\bigskip        

\end{abstract}

\tableofcontents

\newpage
\pagestyle{empty}

\end{titlepage}

\section{Introduction}
\label{sec:intro}

In recent years there has been a growing interest in quantum
field theory over noncommutative spaces\cite{ncspace}, that is spaces where 
the space-time coordinates $x_\mu$, replaced by hermitian operators $\widehat{x}_\mu$, do not
commute:
\begin{equation}
\label{nceqn}
[\widehat{x}_\mu,\widehat{x}_\nu]=i \theta_{\mu \nu}
\end{equation}
Here $\theta$ is a real and antisymmetric object with the dimensions of length-squared and corresponds
to the smallest patch of area in physical space one may 'observe',
 similar to the role $\hbar$ plays in 
$[\widehat{x_{i}},\widehat{p_{j}}]=i \hbar \delta_{ij}$,
 defining the corresponding smallest
patch of phase space in quantum mechanics. 
In this paper we define the energy scale $\Lambda \equiv  {1\over\sqrt{\theta}}$ 
(where $\theta$ is the average magnitude of an element of $\theta_{\mu \nu}$)
which is a more convenient parameterization in constructing an effective
theory at low energies. 
Many researchers set $\theta_{0i}=0$ to avoid problems with unitarity and
causality, but since this is only an issue at energies above $\Lambda$\cite{unitarity}, we
do not use this constraint for the purposes of low-energy phenomenology.
We may view $\theta_{\mu \nu}$ as a ``background $B$-field'' which has
attained a vacuum expectation value, hence appearing in the Lagrangian as
a Lorentz tensor of constants\cite{tensor}. Assuming that 
the components of $\theta_{\mu \nu}$ are constant over cosmological scales,
in any given frame of reference there is a special ``$\nc$ direction'' given
by the vector $\theta^i \equiv \epsilon^{ijk}\theta_{jk}$. Experiments 
sensitive to $\ncg$ will therefore be measuring the components of
 $\overrightarrow{\theta}$, and it is necessary to take into account
the motion of the lab frame in this measurement.  Since $\nc$ effects are measured in powers of
$p^{\mu}   \theta_{\mu \nu} p'^{\nu}$, where $p,p'$ are some momenta involved
in the measurement, it is possible that odd powers of $\theta$ will partially average
to zero if the time scale of the measurement is long enough. Effects of first order in $\theta$
vanish at a symmetric $e^+ e^-$-collider, for example, if the measurement averages 
over the entire $4\pi$ solid angle of decay products. If the data is binned by angle
then it is possible to restore the sensitivity to $\theta$.
In addition to any other averaging process over short time scales, 
terrestrial
 experiments performed over several
days will only be sensitive to the projection of  $\overrightarrow{\theta}$
on the axis of the Earth's rotation. 
 Of course binning the data hourly or
at least by day/night, taking into account the time of year, can partially mitigate this effect.
This axis, as well as the motion of the solar system, galaxy, $\etc$, does not vary over time scales 
relevant to terrestrial experiments.

The basic idea of $\ncg$ is not new and has been known in the 
context of string theory for some time\cite{string}. 
We refer the reader to a few of the many excellent reviews of the mathematics of noncommutative space 
\cite{math1,math2,math3,math4,math5}
for a more rigorous understanding of the present material. 
Noncommuting coordinates are expected on quite general
grounds in any theory that seeks to incorporate gravity into 
a quantum field theory: 
the usual semi-classical argument is that a particle may only be
 localized to within a Planck length $\lambda_P$
without creating a black hole that swallows the particle,
hence $\sum_{i<j}{\Delta x_i \Delta x_j} \geq {\lambda_P}^2$; alternatively,
one is led to think of space as a
noncommutative algebra upon trying to quantize the 
Einstein theory\cite{dubois,doplicher}.

\begin{figure}[t]
\dofig{3.00in}{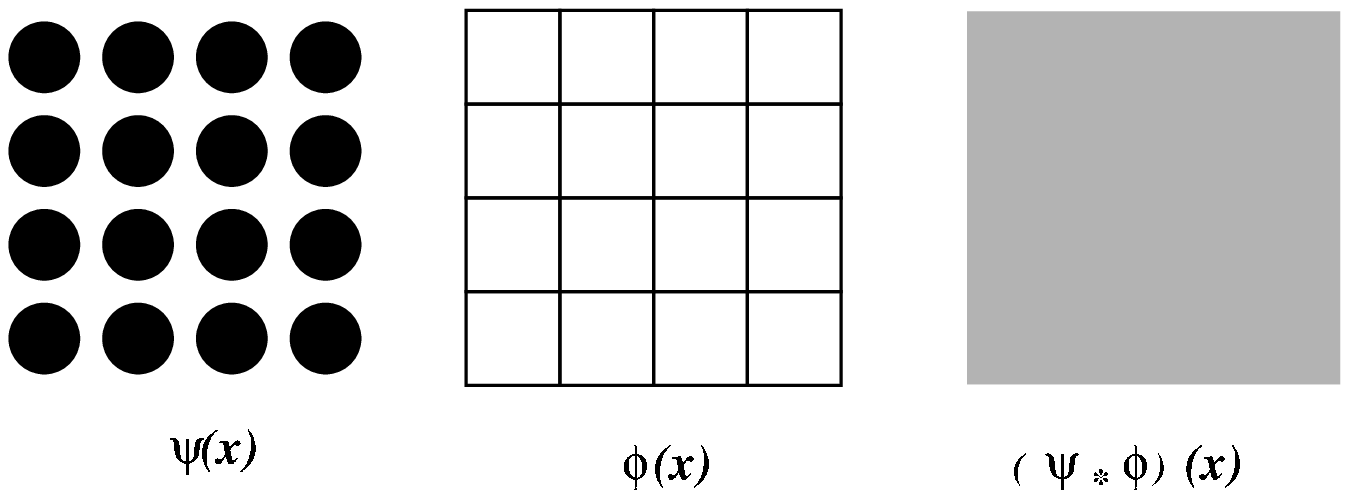}
\caption{An illustration of the star product between two functions. 
    The two scalar functions $\psi$ and $\phi$ are strongly 
    orthogonal($\psi(x) \phi(x) = 0 ~~\forall x$) yet the $\starprod$
is nonzero.
\label{starfig} }
\end{figure}

Much research has already gone into understanding noncommutative quantum 
field theory\cite{ncqft1,ncqft2,ncqft3,ncqft4};
it is equivalent to working with ordinary (commutative) field theory and
replacing the usual
product by the $\starprod$ defined as follows:
\begin{equation}
\label{star}
(f \star g)(x) \equiv e^{i \theta_{\mu \nu} \partial_{\mu}^{y} \partial_{\nu}^{z}}
		f(y) g(z) \mid_{y=z=x}
\end{equation}
With this definition (\ref{nceqn}) holds in function space equipped with a $\starprod$:
\begin{equation}
\label{nceqn2}
[x_{\mu},x_{\nu}]_{\star}=i \theta_{\mu \nu}
\end{equation}
This $\starprod$ intuitively replaces the point-by-point multiplication of two fields
by a sort of 'smeared' product (see Fig. \ref{starfig}). Indeed the concept of 'smearing'
is borne out in more detailed analysis of 1- and 2- point functions\cite{smear}:
spacetime is only well defined down to distances of order $\sqrt{\theta}$ so
functions of spacetime must be appropriately averaged over a neighborhood
of points. In each $(i,j)$ plane, we must replace 
\begin{equation}
\label{avg}
\phi(x_i,x_j) \rightarrow \int dx_i'dx_j' \phi(x') e^{-\frac{(x_i-x_i')^2+(x_j-x_j')^2}{\theta_{ij}}}(\pi \theta_{ij})^{-1} 
\end{equation} 

Examples of theories which have received attention include
scalar field theory\cite{ncqft2,scalar1,scalar2}, 
NCQED (the $\nc$ analog of QED)\cite{ncqed},
as well as noncommutative Yang-Mills\cite{ncym1,ncym2}; perturbation theory in $\theta$ 
is applicable and the theories are renormalizable\cite{renorm1,renorm2}. 
For gauge theories, a suitably adjusted definition of the gauge transformations \cite{wess1,wess2}
permits the construction of $SU(N)$ theories. There has been no work explicitly proving that
fermion representations are consistent with such theories, however we know that the proof must
exist since $\ncg$ is derived from a string theory which of course is self-consistent for all
gauge groups and representations\footnotemark. \footnotetext{We thank B. Zumino for useful discussion on
this point} 

A $\nc$ modification of the Standard Model(SM) is possible as a working field theory, at least 
up to  ${\cal O}(\theta)$. Replacing the ordinary product with the $\starprod$ in the
Lagrangian, the appropriate Feynman rules for this $\nc$ SM (ncSM) follow straightforwardly 
and are reproduced in the Appendix. 
 
Whatever the physics at the Planck scale is like, we expect there
to be some residual effect at low energies beyond that of classical
gravity. If we parameterize this effect as in (\ref{nceqn}),
then low energy physics will receive corrections in powers of the small 
parameter $\theta$. Several papers have addressed how these corrections
may modify observations at an accelerator\cite{hewett}, precision
tests of QED in hydrogen\cite{hydrogen}, and various dipole moments\cite{moments};
in general, if $\Lambda \leq 1~TeV$, there will be some observable effects
in these systems at the next generation of colliders.   
This paper aims to investigate the $\cpviolng$ potential of $\ncg$ in low energy 
phenomenology.

\section{Computing in the Noncommutative Standard Model(ncSM)}
\label{sec:ncqft}

	The method of computing $\nc$ field theory amplitudes is effected by replacing
the ordinary function product with the $\starprod$ in the 
Lagrangian. The theory is otherwise identical to the
commuting one (i.e. the Feynman path integral formulation
provides the usual setting for doing QFT): for example
a Yukawa theory with a scalar $\phi$, Dirac fermion $\psi$, has
the action
\begin{equation}
\label{yukawa}
S=\int{ d^4 x \left( \overline{\psi} i \slashchar{\partial} \psi
+ (\partial \phi)^2 + \lambda \overline{\psi} \star \psi \star \phi \right)}
\end{equation}
(Here we have used the fact that $\int{ dx \xi \star \xi} = \int{ dx \xi \xi}$,
which follows straightforwardly from (\ref{star}) ) 
Gauge interactions likewise generalize from
the standard form; the action for NCQED for example is
\begin{equation}
\label{ncqed}
S=\int{ d^4 x \left(  -\frac{1}{4e^2}F^{\mu \nu} F_{\mu \nu} +  \overline{\psi}  
  i\slashchar{\partial}\psi -  e \overline{\psi} \star \slashchar{A} \star \psi
 - m \overline{\psi} \psi\right)}
\end{equation}
where
\begin{equation}
\label{Fterm}
F_{\mu \nu} \equiv \partial_\mu A_\nu - \partial_\nu A_\mu - i [A_\mu,A_\nu]_\star
\end{equation}
Note the extra term in the field strength which is absent in ordinary QED; this
nonlinearity gives NCQED  a NonAbelian-like structure. There will be, for example,
3- and 4-point photon self-couplings at tree level (see Appendix).  

In momentum space the $\starprod$ becomes a momentum-dependent phase factor which
means that the theory effectively contains an infinite number of derivative interactions
suppressed by powers of $\theta$. This directly exhibits the nonlocal character
of $\ncg$. From (\ref{yukawa}) and (\ref{ncqed}) we can derive
the action for the $\nc$ version of the Standard Model (ncSM).  We present its content
as the list of Feynman rules in the Appendix. 

A central feature of computations in the SM is the presence of divergences and the
need to absorb them into counterterms. The ncSM is similar    in this respect, yet
it is necessary to renormalize carefully: if one simply uses dimensional 
regularization and sums virtual energies to infinity, bizarre infrared singularities
appear in the theory which are difficult to handle\cite{ncqft2}. 
To illustrate, consider the loop integral
\begin{equation}
\label{qloop}
\int{d^d k \frac{e^{i k^\mu \theta_{\mu \nu} p^\nu}}{(k^2-m^2)^2}}
\end{equation}
which is finite for $\abs{\theta \cdot p} \neq 0$ but is logarithmically
divergent if $\abs{\theta} = 0$.
Explicitly, we Wick-rotate (\ref{qloop}), 
introduce the Schwinger parameters\cite{schwinger}, integrate
over momenta, and obtain
\begin{equation}
\label{loop2}
\int{dS~S^{1-{d\over2} }~ e^{-\frac{1}{4}(\theta \cdot p)^2 S^{-1} - m^2 S}}
\end{equation}
If we take  $\abs{\theta} = 0$ now, dimensional regularization  
 gives the usual $\Gamma[1- {d\over2}]$ which we would absorb into a counterterm
of the theory. However for small finite values of $\abs{\theta \cdot p}$  we
get an approximation of the integral  (\ref{loop2}) in four dimensions: 
\begin{equation}
\label{loop3}
 \int{d^4 k \frac{e^{i k^\mu \theta_{\mu \nu} p^\nu}}{(k^2-m^2)^2}} \approx
 ln(m^2\abs{\theta \cdot p}^2) \left( 1 + m^2\abs{\theta \cdot p}^2 \right)
\end{equation}
There is a $ln(\abs{\theta})$ divergence as  $\abs{\theta} \to 0$ which
is expected since in this limit the theory tends to the commutative one 
and reproduces the  $\Gamma[1- {d\over2}]$ divergence mentioned above. This is
formally correct, however the theory 
in this limit is awkward to work with since some contributions will diverge as
 $\abs{\theta} \to 0$ and must produce final results
 such as scattering amplitudes which are finite.
For the computational purposes of this paper, in which $\eg$ ${m_W}^2\abs{\theta} \ll 1$,
 it is more convenient to regularize
with a Pauli-Vilars regulator with mass $M$. 
Then  (\ref{loop2}) becomes
\begin{equation}
\label{loop4}
\int{dS~S^{1-{d\over2} }~ e^{-(M^{-2}+\frac{1}{4}(\theta \cdot p)^2) S^{-1} - m^2 S}}
\end{equation}
Taking
the limit  $\abs{\theta} \to 0$ now gives
\begin{equation}
\label{loop5}
 \int{d^d k \frac{e^{i k^\mu \theta_{\mu \nu} p^\nu}}{(k^2-m^2)^2}} \approx
 ln(\frac{m^2}{M^2} + m^2 \abs{\theta \cdot p}^2) +
  m^2 \abs{\theta \cdot p}^2 ln(\frac{m^2}{M^2} + m^2 \abs{\theta \cdot p}^2)
\end{equation}
Note that in the limit  $\abs{\theta} \to 0$ the second term vanishes
while the first term reproduces the 
ordinary(commutative) loop integral divergence. We subtract this into a counterterm,  
while the remaining piece gives a small
correction to the commutative theory of 
${\cal O}(x~ln(x))$ where $x\equiv \abs{m~\theta \cdot p}^2$. Renormalizing in this manner
guarantees sensible results.

\section{CP Violation in the ncSM}
\label{sec:cpviol}
	In the SM, there are only two sources of $\cpviol$: the 
irremovable phases in the CKM matrix and the $\Theta F\tilde{F}$ term
in the strong interaction Lagrangian(the coefficient $\Theta$ 
has to be miniscule to avoid contradicting experiment\cite{strongCP}). 

	In the ncSM, there is an additional source of $\cpviol$:
the parameter $\theta$ itself is the $\cpviolng$ object, which is
apparent from the NCQED action (\ref{ncqed}) considering the
transformation of $A_\mu$ and $\partial_\mu$ under $C$ and $P$
and assuming $CPT$ invariance\cite{cpt}. Physically speaking, an area of ${\cal O}(\theta)$ represents
a ``black box'' in which some or all spacetime coordinates become ambiguous, which in turn leads
to an ambiguity between particle and antiparticle. More detailed work reveals that $\theta$ is 
in fact proportional to the size of an effective particle dipole moment\cite{dipole}. Therefore
$\ncg$ can actually {\it explain} the origin of $\cpviol$. 
At the field theory level, it is the momentum-dependent phase factor appearing in the $\nc$
theory which gives $\cpviol$. For example, the ncSM W-quark-quark $SU(2)$ vertex
in the flavor basis is
\begin{equation}
\label{wqq}
{\cal L}_{Wqq} =  \overline{u(p)} \gamma^\mu (1-\gamma_5) e^{ip\cdot \theta \cdot p'}~d(p')~W_\mu  
\end{equation}
Once we perform rotations on the quark fields to diagonalize the Yukawa interactions, 
$\ie$ $u_L \to U u_L$ and $d_L \to V d_L$, the above becomes
\begin{equation}
\label{wqqmass}
{\cal L}_{Wqq} =  \overline{u(p)} \gamma^\mu (1-\gamma_5) 
e^{ip\cdot \theta \cdot p'} U^\dagger V ~d(p')~W_\mu  
\end{equation}
Even if $U^\dagger V$ is purely real, there will be some nonzero phases $e^{ip\cdot \theta \cdot p'}$ 
in the Lagrangian whose magnitudes increase as the momentum flow in the process increases. Of course,
the above phase factor has no effect at tree-level (suitably redefining all the fields) but will
affect results at 1-loop and beyond. 

Experimentally, the signal for $\ncg$ here is a momentum-dependent
 CKM matrix (ncCKM) which we define as follows:
\begin{equation}
\label{ncCKM}
\overline{V}(p,p') \equiv  
	  \left( \begin{array}{ccc} 
	1- \lambda^2/2 + ix_{ud} & \lambda  + i\lambda x_{us}  &
         A \lambda^3 (\rho-i\eta) + iA\lambda^3 \rho x_{ub} \\
	-\lambda -  i \lambda x_{cd}& 1- \lambda^2/2 + ix_{cs} &
         A \lambda^2 +  iA \lambda^2 x_{cb} \\
	 A \lambda^3 (1-\rho-i\eta)+  i A \lambda^3 \rho x_{td}  & -A \lambda^2 - A \lambda^2 ix_{ts} &
          1 +  ix_{tb}\\ \end{array} \right)
\end{equation}
where $x_{ab} \equiv {p_a}^\mu \theta_{\mu \nu} {p'_b}^\nu$ for quarks $a,b$. 
This matrix is an approximation of the exact ncSM in the perturbative limit
where we expand
$ e^{ip\cdot \theta \cdot p'} \approx 1 + ip\cdot \theta \cdot p'$ \footnotemark
\footnotetext {We thank D.A. Demir for help in clarifying the notation}.
In the limit $\theta \to 0$, the $x_{ab}$ all go to zero and $\overline{V}$
becomes the CKM matrix $V$ in the Wolfenstein parameterization\cite{wolf} in terms of the
small number $\lambda \approx 0.22$. 
Note that $\overline{V}$ is not guaranteed to be unitary, since, in contrast to the SM CKM matrix,
 $\overline{V}$ is not a collection of
derived constants: a given matrix element will
attain different values depending on the process it is describing. As an example,
suppose we measure a non-zero $\tau$-polarization asymmetry in 
$t \to b \tau^+ \nu$\cite{taupol}; this puts
a constraint on the value of 
$\Im(\overline{V}_{tb})$ {\it at the energy scale $\mu \approx m_t$}\footnotemark.
\footnotetext{Actually, there is a lot of uncertainty in this measurement, including
the values of the $MNS$ matrix\cite{mns}, so measuring the phase in practice is not straightforward.} 
 We can
get another constraint on $\Im(\overline{V}_{tb})$ through a $B^0-\overline{B}^0$ 
oscillation experiment, but 
we must take into consideration that this is a measurement at the energy scale $\mu \approx m_b$.
In the former process we would find (for $\eta=0$) 
 $\Im(\overline{V}_{tb}) \approx {\cal O}({m_t}^2\abs{\theta})$ whereas in
the latter it would be  ${\cal O}(m_t m_b\abs{\theta})$, so these phases differ by
a factor of $m_t/m_b \approx 30$. Therefore we expect the phenomenology of $\overline{V}$ 
to be rather different from that of the SM. 
In addition to $\cpviol$ from the weak interaction (in $\overline{V}$),
 there will also be $\cpviol$ from the strong and electromagnetic interactions
(since there are phases entering any vertex with three (or more) fields (see Appendix)).
We now turn to the phenomenological implications of these.
 
\subsection{CP Violating Observables}
\label{sub:cpviolob}
\subsubsection{$\epsilon_K$}

\begin{figure}[t]
\dofig{3.00in}{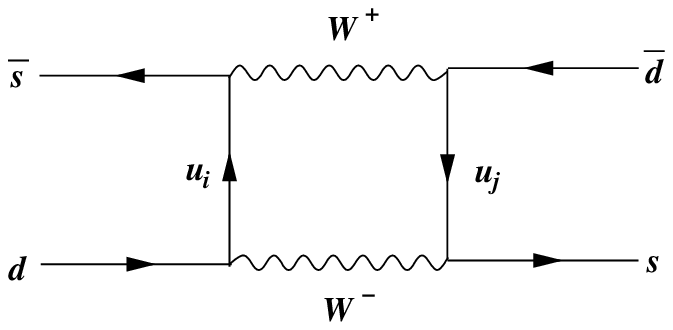}
\caption{The box graph for $K^0$-$\overline{K^0}$ mixing
in the Standard Model with exchange of virtual $W$-bosons
and up-type quarks from the $i,j$ generations. 
\label{boxfig}} \end{figure}

The $\cpviolng$ observable of choice in the $K^0$-meson system is $\epsilon_K$ which is
directly proportional to the imaginary part $\Im(M_{12})$ of the box graph (see Fig. \ref{boxfig}):
\begin{equation}
\label{epsilonk}
\epsilon_K \equiv \frac{\Im(M_{12})}{\Delta m}
\end{equation}
The mass splitting $\Delta m$ between the long- and short-lived $K^0$ eigenstates
\newline
 is $\Delta m \approx 3.5 \cdot 10^{-15}~GeV$\cite{pdg}. We can rewrite
\begin{equation}
\label{m12}
\Im(M_{12}) = \frac{{G_F}^2 {m_W}^2 {f_K}^2 B_K m_K} {12 \pi^2} \Im(loop)
\end{equation}
in terms of the decay constants $f_K,B_K$, and the loop factor. In the SM, the loop factor is
\begin{equation}
\label{loop}
\Im(loop) \approx \Im({\lambda_c}^2 f(m_c) + {\lambda_t}^2 f(m_t) +  
     \lambda_c  \lambda_t f(m_c,m_t))
\end{equation}
\newline
where $\lambda_q \equiv V_{qd} V_{qs}^*$ and $f(x)$ is a loop function (see Appendix).
In the SM, both charm and top quarks contribute roughly equally to the imaginary part of
the loop, and the measured value for $\epsilon_K$ puts a constraint on the
parameters $\rho,\eta$ of the CKM matrix. However, in the ncSM we must replace the
 entire loop since the
momentum-dependent phases in $\overline{V}$ change how the loop integral behaves.
Note the charm quark will dominate the imaginary part of the graph because 
the phase of the product  $(\overline{V}_{ts}^*\overline{V}_{td})^2$  is a factor of $\lambda^8$ suppressed
relative to the phase of  $(\overline{V}_{cs}^*\overline{V}_{cd})^2$ (see (\ref{ncCKM})). 
We record the evaluation of the loop integral in the Appendix.

If the kaons used in the measurement emerge from a beam with an average
velocity $\beta \equiv {v \over c}$ in the lab frame, we must average over
the motion of the internal constituents of the kaon, since the entire 
$\nc$ effect is proportional to $p \cdot \theta \cdot p'$, where
$p,p'$ are the momenta of the constituents. We assume
that these momenta have random orientation in the rest frame of the kaon,
subject to $p + p' = (m_K, 0,0,0)$. The average over these internal
momenta produces a result which is proportional to the velocity of the kaon
in the lab frame: $\langle p \cdot \theta \cdot p' \rangle \approx 
\abs{\theta} \beta \gamma~ m_K^2$. Therefore it is important that
the  $\beta \gamma$ of the beam not be so small as to wash out the signal.
Recent determinations of $\epsilon_K$ use a reasonable  $\beta \gamma$\cite{cplear}, so
we do not concern ourselves further with this caveat. Experiments
at an $e^+e^-$ collider ($\eg$\cite{tac,daphne}) where the center of mass is stationary in the lab frame
 should, however, see no signal for $\epsilon_K$ since 
$\langle \overrightarrow{\beta}\gamma \rangle = 0$.
 As we mentioned in the 
Introduction, the data may be sensitive to the time of day. If there is a component
of $\overrightarrow{\theta}$ along the axis of the Earth, then given
enough statistics there should
be a ``day/night effect'' for $\epsilon_K$ which, as far as we know, no 
experiment has looked for. 

\begin{figure}[t]
\dofig{3.00in}{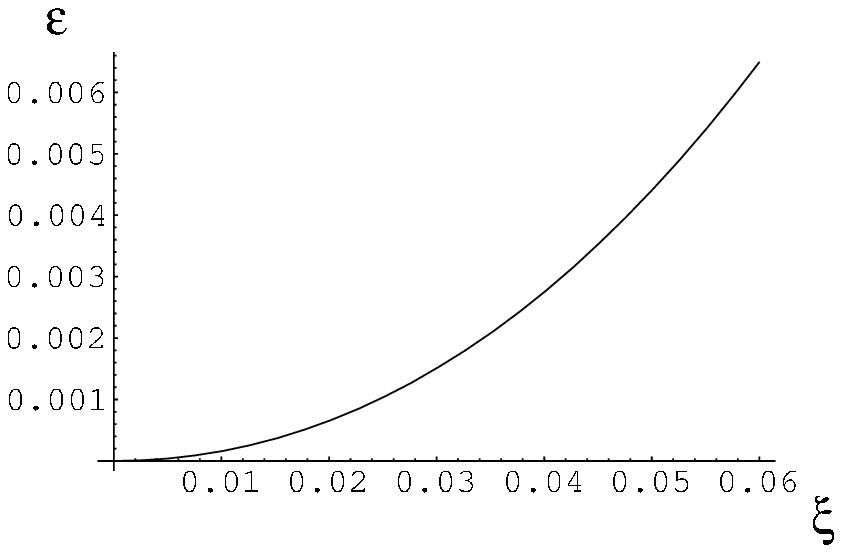}
\caption{Variation of $\epsilon$ with $\xi \equiv m_W / \Lambda$.
Here $\eta=0$ so all $\cpviol$ is from $\ncg$. 
\label{epsilonfig} }

\end{figure}

In the case $\eta=0$ (so the phase from $\overline{V}$ is due entirely to $\ncg$),  we obtain
\begin{equation}
\label{ekresult}
\begin{array}{l}
\epsilon_K \approx \frac{{G_F}^2 {m_W}^2 {f_K}^2 B_K m_K} {12 \pi^2 \Delta m} 
	\lambda^2 \frac{m_K}{m_W} \xi^2 \\
\\
\xi \equiv \frac{m_W}{\Lambda}
\end{array}
\end{equation}
Using $G_F = 1.166 \cdot 10^{-5}~GeV^{-2}, m_W = 80.4~GeV, 
 f_K = 0.16~GeV, m_K = 0.498~GeV, B_K = 0.70 \pm 0.2, \rho = 0.3 \pm 0.2$, and  the latest measurement of
 $\epsilon_K \approx (2.280 \pm 0.013)\cdot 10^{-3}$\cite{pdg}, this 
implies $\xi \approx (4\pm2)\cdot 10^{-2}$ (see Fig. \ref{epsilonfig}); in this scenario spacetime
becomes effectively $\nc$ at energies above $\approx 2~TeV$. 

\subsubsection{$\epsilon'/\epsilon$}
Direct $\cpviol$ is measurable in the neutral kaon system as a difference
between the rates at which $K_{L,S}$ decay into $I=0,2$ states of pions:
\begin{equation}
\epsilon' \equiv \frac{\langle 2 \mid T \mid K_L \rangle \langle 0 \mid T \mid K_S \rangle 
- \langle 2 \mid T \mid K_S \rangle \langle 0 \mid T \mid K_L \rangle}
	{\sqrt{2}\langle 0 \mid T \mid K_S \rangle^2 }
\end{equation} 
Then the ratio of direct to indirect $\cpviol$ is
\begin{equation}
\frac{\epsilon'}{\epsilon} = \frac{1}{\sqrt{2}} \left( 
	\frac{\langle 2 \mid T \mid K_L \rangle}{\langle 0 \mid T \mid K_L \rangle}
   -  \frac{\langle 2 \mid T \mid K_S \rangle}{\langle 0 \mid T \mid K_S \rangle}  \right)
\end{equation}
The theoretical computation of this ratio is a challenge in the SM not only 
because the perturbative description of the strong interaction is not 
reliable at low energies but also because it is proportional to a difference between
two nearly equal contributions, enhancing the theoretical error\cite{epprime}.
The most naive way to estimate 
$\frac{\epsilon'}{\epsilon}$ employs the so-called 
vacuum-saturation-approximation (VSA) which is based on the factorization of
four-quark operators into products of currents and the use of the vacuum as
an intermediate state (for more details see\cite{vac}). The estimate is
\begin{equation}
\frac{\epsilon'}{\epsilon} \approx (0.8 \pm 0.5) \cdot 10^{-3} 
	\left( \frac{\Im(\lambda_t)}{10^{-4}} \right) 
\end{equation}
where in the SM $\lambda_t$ represents the $\cpviolng$ phases from the CKM
matrix,  $\lambda_t = A^2 \lambda^5 \eta \approx 1.3 \cdot 10^{-4}$. The
experiments measure $\frac{\epsilon'}{\epsilon} = (1.92 \pm 0.46) \cdot 10^{-3}$
which does not closely match the VSA number, but it is possible to use more
elaborate models that agree closely with the measured value\cite{epprime}.

In the ncSM it is no less difficult to compute $\frac{\epsilon'}{\epsilon}$; in 
particular, the extra phases from 
\newline
 $\ncg$ will become involved in the complicated
nonperturbative quark-gluon dynamics. The best estimate we can make here is
(see Appendix \ref{app:eps}) 
\begin{equation}
\label{penguin}
\Im(\lambda_t) \approx 2\frac{m_K}{m_W} \xi^2~ln(\frac{m_W}{\xi m_K})
\end{equation}
For $\xi \approx 0.04$, we get roughly the same VSA value as in the SM.

\subsubsection{$sin2\beta$ and the unitarity triangle}
\label{subsub:s2b}

The only $\cpviolng$ observation from the $B$-system to date,
the asymmetry in the decay products of $B^0 \to J/\psi {K_s}^0$ 
\cite{psikshort1,psikshort2,psikshort3,psikshort4},
is a measurement in the SM
of a combination of CKM elements called $sin2\beta$:
\begin{equation}
\label{sin2beta}
sin2\beta \equiv  \Im \left( - \left[\frac{V_{tb}^* V_{td}}{V_{tb}V_{td}^*}\right]
			 \left[\frac{V_{cs}^* V_{cb}}{V_{cs}V_{cb}^*}\right]
			 \left[\frac{V_{cd}^* V_{cs}}{V_{cd}V_{cs}^*}\right]
\right)
\end{equation}
where the first bracketed factor is from ${B_d}^0-\overline{B_d}^0$ mixing,
the second from the observed decay asymmetry, and the third from 
 ${K}^0-\overline{K}^0$ mixing. In the Wolfenstein parameterization,
\begin{equation}
sin2\beta \approx \frac{2 \eta (1-\rho)}{\eta^2 + (1-\rho)^2}
\end{equation}
which, for $(\rho,\eta) \approx (0.2,0.3)$, corresponding to a point
in the center of the allowed region of the $\rho-\eta$ plane\cite{rhoetaplane}
implies $sin2\beta \approx 0.7$. The most recent experimental world average
for this quantity is $\approx 0.49 \pm 0.23$\cite{sin2b}.

In the ncSM the corresponding quantity is (\ref{sin2beta}) with 
each matrix element $V_{ij}$ replaced by $\overline{V}_{ij}$ extracted
from the relevant process:
\begin{equation}
\label{sin2beta2}
sin2\beta \to  \Im \left( - \left[\frac{\overline{V}_{tb}^* 
		\overline{V}_{td}}{\overline{V}_{tb}\overline{V}_{td}^*}\right]
		\left[\frac{\overline{V}_{cs}^* 
		\overline{V}_{cb}}{\overline{V}_{cs}\overline{V}_{cb}^*}\right]
		\left[\frac{\overline{V}_{cd}^* 
		\overline{V}_{cs}}{\overline{V}_{cd}\overline{V}_{cs}^*}\right]
\right)
\end{equation}
Of course experiments don't measure the precise value of a given $\overline{V}_{ij}$
but rather some combination of them integrated over internal momenta. If we again
consider the scenario where $\eta = 0$ then
the imaginary parts of these quantities increase roughly proportionally to the momenta
 involved and we expect the first bracketed term in (\ref{sin2beta2}) to dominate since 
the size of the momenta involved in ${B_d}^0-\overline{B_d}^0$ mixing exceeds that
of $B^0$-decay or ${K}^0-\overline{K}^0$ mixing, $\ie$ 
$m_b m_t \theta \gg m_b^2 \theta, m_K m_t \theta$.  We therefore set the
second and third brackets to unity, obtaining
\begin{equation}
\label{sin2beta3}
sin2\beta \approx \Im \left(\frac{1-i x_{tb}}{1+ix_{tb}} \right) \approx \frac{m_b}{m_W}\xi^2 
\end{equation}
If we use the measurement of $\epsilon_K$ to fix $\xi \approx 10^{-2}$, then the ncSM predicts
$sin2\beta \approx 0$ which is not excluded by experiment.
The motion of the quarks inside the $B$-meson moreover partially washes out the signal
(see previous discussion for kaons)
as the asymmetry of the $e^+ e^-$ collider gives
the  ${B_d}^0-\overline{B_d}^0$ center-of-mass only a modest boost of $\beta \gamma \approx 0.6$
in the lab frame. We conclude that this model predicts that current $B$-physics experiments
should see a value of $sin2\beta$ which is consistent with zero. 

\begin{figure}[t]
\dofig{3.00in}{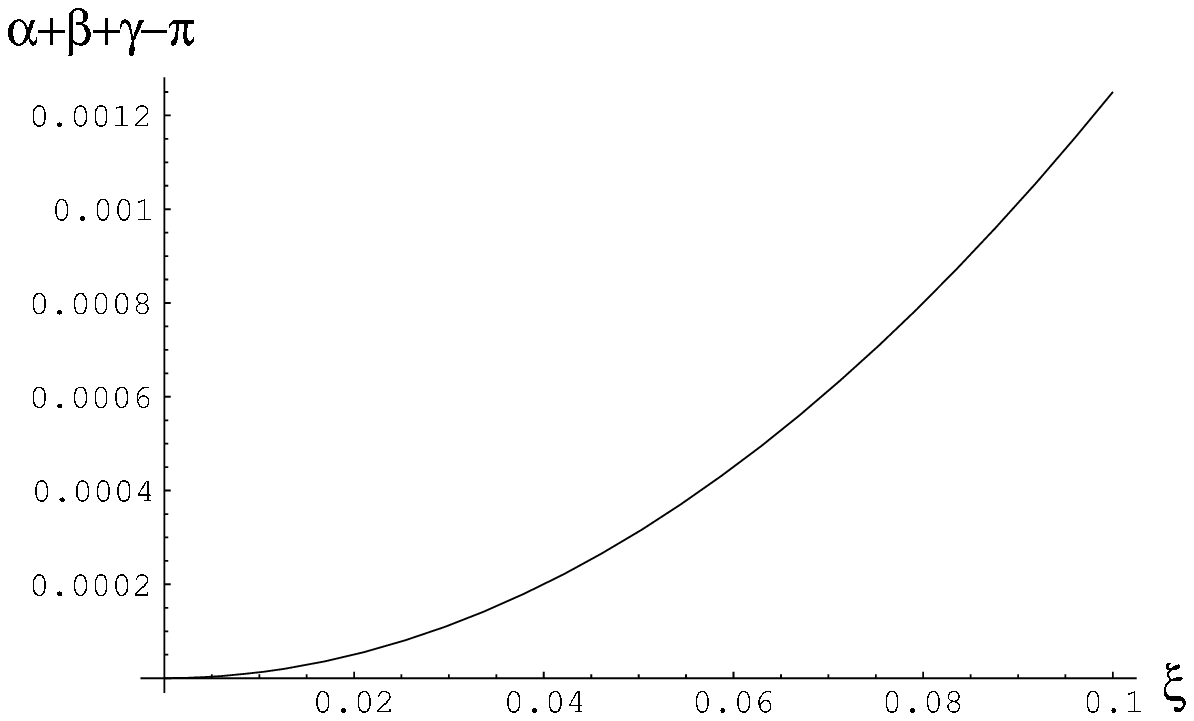}
\caption{Plot of the difference $\alpha+\beta+\gamma-\pi$ in the ncSM
 illustrating that
the unitarity triangle does not close exactly. \label{anglefig}
}
\end{figure}

The other two $\cpviolng$ observables commonly defined in $B$-physics are
$\alpha$ and $\gamma$:
\begin{equation}
\label{angles1}
\alpha \equiv arg\left( -\frac{V_{tb}^* V_{td}}{V_{ud}V_{ub}^*} \right)
~~~~~~~~~~~~~~~~~~~~~
\gamma \equiv arg\left( -\frac{V_{cd}^* V_{cb}}{V_{ud}V_{ub}^*} \right)
\end{equation}
where $V_{tb}^* V_{td}$ can be extracted from  ${B_d}^0-\overline{B_d}^0$ mixing and
$V_{ud}V_{ub}^*$, $V_{cd}^* V_{cb}$  from  neutral and charged $B$-decays such as 
$B^0 \to \pi \pi$ and $B^+ \to \pi^+ K^0$, for example. In the SM $\alpha + \beta + \gamma = \pi$ 
because the CKM matrix $V$ is unitary. The ncSM matrix $\overline{V}$ is
not unitary (see Section \ref{sec:cpviol}), so we expect 
 $\alpha + \beta + \gamma \ne \pi$ as these ``angles'' are defined (by
$\overline{V}$ replacing $V$ in (\ref{angles1}) above). For $\eta=0$
the parameters $\alpha, \beta, \gamma$ in the ncSM assume the following
form:
\begin{equation}
\label{angles2}
\begin{array}{l}
\alpha \approx tan^{-1} \left( \frac{m_b}{m_W} \xi^2 \right) \\
\beta \approx tan^{-1} \left( \frac{m_b+m_K}{m_W} \xi^2 \right) \\
\gamma \approx  \pi - tan^{-1} \left( \frac{m_K}{m_W} \xi^2 \right) \\
\end{array}
\end{equation}
In Figure \ref{anglefig} we plot the sum $\alpha + \beta + \gamma$. The
angles essentially add up to $\pi$ in the same range of $\xi$ which is
required by the $\epsilon_K$-constraint. If all the matrix elements of
 $\overline{V}$ could be measured at the same energy then the unitarity
triangle would close exactly. The small deviation from exact closure is
${\cal O}(\frac{m_b}{m_W} \xi^2)$ and represents the fact that the angles
as defined in (\ref{sin2beta2}) and (\ref{angles1}) are a combination of
amplitudes measured at different energies: $\mu \sim m_t$ (for  ${B_d}^0-\overline{B_d}^0$ mixing) and
and  $\mu \sim m_b$ (for $B$ decays). 

\subsubsection{Electric Dipole Moments}

Nonzero values of the electric dipole moments ($\edm$s) of the elementary fermions 
necessarily violate $T$, and hence $CP$ (assuming the $CPT$ theorem). This follows from
the observation that a dipole moment $\overrightarrow{D}$ is a directional quantity,
so for an elementary particle it must transform like the spin $\overrightarrow{J}$, the only 
available directional quantum number. The interation with an external
electric field $\overrightarrow{E}$ is
 $\overrightarrow{J}\cdot \overrightarrow{E}$ which is therefore $CP$-odd. 
The presence of an $\edm$ for a particle $\psi$ implies an interaction 
with the electromagnetic field strength $F^{\mu \nu}$ in the
Lagrangian of the form 
$$
O_{edm} = - (i/2)\overline{\psi}  \gamma_5 \sigma_{\mu \nu} \psi F^{\mu \nu}
$$
In the SM this operator is absent at tree level and even at one loop
due to a cancellation of the CKM phases. For the electron, moreover,
 the $\edm$ ($d_e$)
 vanishes at two loops and the 
three-loop prediction is miniscule, of order $10^{-50}e~cm$\cite{donoghue}. For the 
neutron $\edm$ ($d_n$), gluon interactions can give rise to a two-loop contribution which
is ${\cal O}(10^{-33}) e~cm$. Upper limits from experiments exist:
$d_e \leq 4.3 \cdot 10^{-27} e~cm$\cite{commins},
$d_n \leq 6.3 \cdot 10^{-26} e~cm$\cite{harris}.

Since the SM predictions of $\edm$s are almost zero, we might expect that new 
sources of $\cpviolng$ physics from $\ncg$ would be observable. The 
noncommutative geometry  provides
in addition a simple explanation for this type of 
$\cpviol$: the directional sense of $\overrightarrow{D}$ derives from
the different amounts of noncommutivity in different directions 
($\ie$ $D_i \propto \epsilon_{ijk}\theta^{jk}$) and the 
size of the $\edm$, classically proportional to the spatial extent of a charge
distribution, is likewise in $\ncg$ proportional to  $\abs{\theta}$, the
inherent ``uncertainty'' of space.
The effects of $\ncg$ will be proportional to the typical momentum
involved, which for an electron $\edm$ observation is $\sim keV$. A detailed
analysis of the size of the $\edm$ appears in \cite{moments}, but a simple
estimate of the expected dipole moment is
\begin{equation}
\label{edmeqn}
d_e \sim e \abs{p \theta} \approx 10^{-20} (\frac{p_e}{1~keV})\xi^2 ~ e\cdot cm
\end{equation} 
which gives an apparently strong upper bound: $\xi < 10^{-3}$. Although the 
phenomenologically interesting values of 
$\xi$ from the $K$-sector is well above this bound, we cannot exclude the
possibility that the actual $\edm$ is much smaller than the above naive
estimate, a situation which can arise in supersymmetric models \cite{susyedm1,
susyedm2}.

\section{Constraints from $\bf {g-2}$ of the Muon}

Since $\nc$ effects are proportional to momentum, we might expect an even
stronger constraint by 
considering the muon $\edm$ in an experiment using relativistic muons,
however the experimental bound here is weaker: 
$d_{\mu} < 1.05 \cdot 10^{-18} e~cm$\cite{muonedm}.

The recent measurement of the anomalous magnetic moment of the muon\cite{bnl}, $a_{\mu}$, 
although not a $\cpviolng$ observable,
does however provide an interesting constraint on the ncSM. Experiments dedicated to
 $a_{\mu}$ have undergone continual refinement (for history
and experimental details, see\cite{muonedm},\cite{muonold})
to the point where   $a_{\mu}$ is now very precisely known:
\begin{equation}
a_{\mu}^{expt} = 11659202(14) \cdot 10^{-10}
\end{equation}
The experimental technique employs muons trapped in a storage ring. A 
uniform magnetic field $B$ is applied perpendicular to the orbit of the 
muons; hence the muon spin will precess. The signal is a discrepancy between
the observed precession and cyclotron frequencies.

Precession of the muon spin is determined indirectly from the decay
$\mu \to e~{\overline \nu}_e ~ \nu_\mu$. Electrons emerge from
the decay vertex with a characteristic angular distribution which
in the SM has the following form in the rest frame of the muon:
\begin{equation}
\label{asymm}
dP(y,\phi) = n(y) (1 + A(y) cos(\phi))dy d(cos(\phi))
\end{equation}
where $\phi$ is the angle between the momentum of the electron
and the spin of the muon, $y = 2 p_e/m_\mu$ measures the fraction of the maximum 
available energy which the electron carries, and $n(y),A(y)$ are 
particular functions which peak at $y=1$. The detectors (positioned 
along the perimeter of the ring) accept the passage of only the highest
energy electrons in order to maximize the angular asymmetry 
in (\ref{asymm}). In this way, the electron count rate is
modulated at the frequency $a_\mu e B/(2\pi m c)$. 

\begin{figure}[t]
\dofig{4.00in}{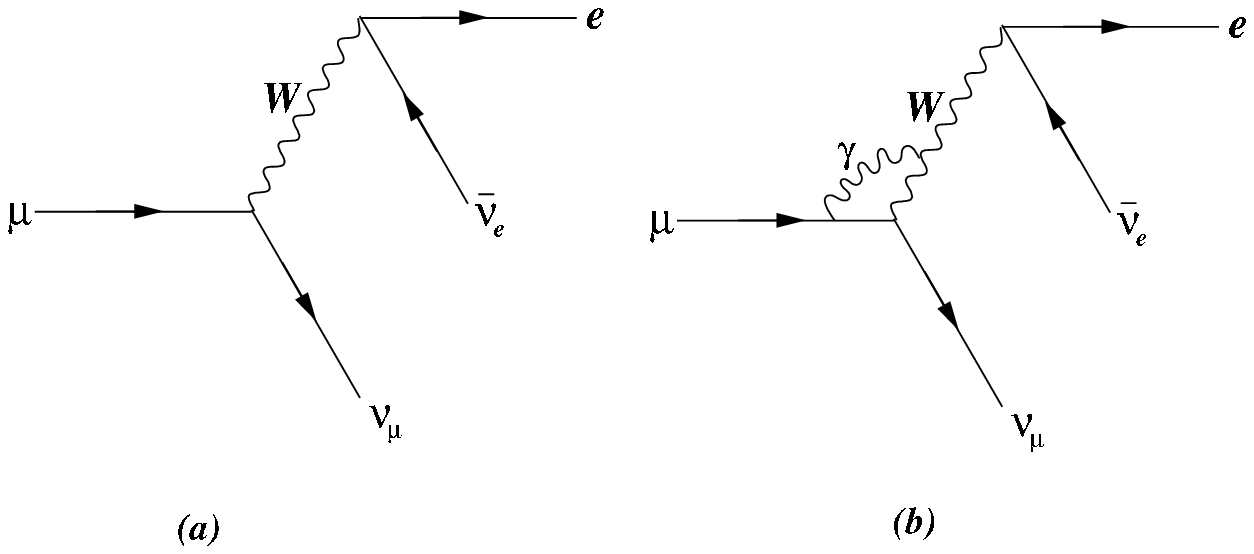}
\caption{Contributions to muon decay (a) SM tree level
(b) ncSM graph which upsets the electron's angular distribution 
 \label{muonfig} }
\end{figure}

Although $a_\mu$ does receive a sizable contribution from $\ncg$, 
it is a {\it constant} contribution\cite{moments}, $\ie$ the
interaction with the external magnetic field $\Delta E \sim B_i \theta_{jk} \epsilon^{ijk}$
is independent of the muon spin, and therefore the
experiment described above is not sensitive to this perturbation of $a_\mu$. 
The effect of $\ncg$ on this measurement does however enter in the manner in
which the muon spin is measured in its decay. Specifically, the electron decay
distribution (\ref{asymm}) has a slightly different angular dependence 
due to the departure of the ncSM from the standard V-A theory of the weak
interactions (see Figure \ref{muonfig}). The electron distribution $dP'$ in 
the ncSM differs from the SM (we reserve the details for a future publication):
\begin{equation}
\label{asymm2}
\begin{array}{ll}
dP'(y,\phi) & \approx n(y) \left( 1 + A(y)(\overrightarrow{p_e} \cdot \overrightarrow{s}_\mu)
		+ f(y)(\hat{p}_e \cdot \theta \cdot \overrightarrow{s}_\mu)
		(\overrightarrow{p_e} \cdot \overrightarrow{s}_e)+ \cdot \cdot \cdot \right) dy d\Omega \\

& \to n(1) \left( 1 + A(1)cos(\phi)
		+ f(1)sin(2\phi)\abs{\theta} + \cdot \cdot \cdot \right) dy d\Omega \\
\\
{\rm where} &  \abs{{f(1) \over A(1)}} \approx \frac{\alpha}{16\pi^2} \frac{p_\mu}{m_W} \xi  \\
\end{array}
\end{equation}
The effect of $\ncg$ is greater than one would naively expect as, for reasons of efficiency, 
the muons are stored at highly relativistic energies: $p_\mu \approx 3~GeV$. Hence the
ratio $ \abs{{f(1) \over A(1)}} \approx 10^{-6} \xi$. However, the frequency is measured
over many cycles and a more conservative estimate of the effective size of the $\nc$ term
is closer to $ (10^{-7} ~to~10^{-8}) \xi$ 
The angular distribution is therefore
not a pure $cos(\phi)$ and we expect the measurement of the precession frequency to
differ from the SM prediction at the level of 1 part in $10^8$. 

Currently, the discrepancy between the measured value of $a_\mu$ and the SM prediction is
\begin{equation}
a_\mu^{expt} - a_\mu^{SM} =  43(16) \cdot 10^{-10}
\end{equation}
which imposes the constraint $\xi \leq 5 \cdot 10^{-2} $. This bound accomodates the 
values of $\xi$ inferred from
$\cpviolng$ observables in section \ref{sub:cpviolob}. 
We expect the value of $\xi$ determined from a $g-2$ experiment to be smaller than
that from a $K$ or $B$-physics experiment since the circulation of the
muons at their cyclotron frequency introduces an additional averaging of the
components of $\overrightarrow{\theta}$. For a storage ring located at an Earth
lattitude of $\psi$ degrees, there will be a $sin(\psi)$ suppression factor.

\section{Conclusions}

The Standard Model(SM) is a highly successful  effective
theory for energies below the weak scale $\sim 100~GeV$, but it must
eventually give way to a description of nature that includes gravity.
Noncommutative geometry is one candidate for such a description,
exhibiting some features of gravity such as nonlocality and 
space-time uncertainty. 

In this paper we have considered the potential effects at
low eneriges of a 
noncommutative geometry which sets in at some high scale $\Lambda$. Remarkably,
for $\Lambda$ in the $TeV$- range, $\nc$ contributions to
$\cpviolng$ observables such as $\epsilon_K$ and
$\epsilon'/\epsilon$ are competitive with
the SM contributions, whereas $sin2\beta \approx 0$. If  $\Lambda \sim 2~TeV$, the
predictions of these observables from $\ncg$ is consistent with
data. Moreover the recent $2.6~\sigma$ deviation between
the SM prediction of $(g-2)$ of the muon and data is 
explained in the $\nc$ scenario for this same value
of $\Lambda$. These perturbative results in terms of the
small parameter $\xi \equiv m_W/\Lambda$
are encouraging, but more work is needed in the treatment
of the full, nonperturbative theory.
Nonetheless, noncommutativity of the space-time
coordinates offers a more physical interpretation
of $\cpviol$ which, if correct, suggests interesting
physics at $TeV$ energies.

\section*{Acknowledgements}
We thank Bruno Zumino and Sheikh-Jabbari for much useful discussion.
 This work was supported by the Director, Office of Science, Office
of Basic Energy Services, of the U.S. Department of Energy under
Contract DE-AC03-76SF0098.

\pagebreak

\appendix\section{Feynman Rules in the NCSM}
\label{app:feyn}
\begin{figure}[hb]
\dofig{7.00in}{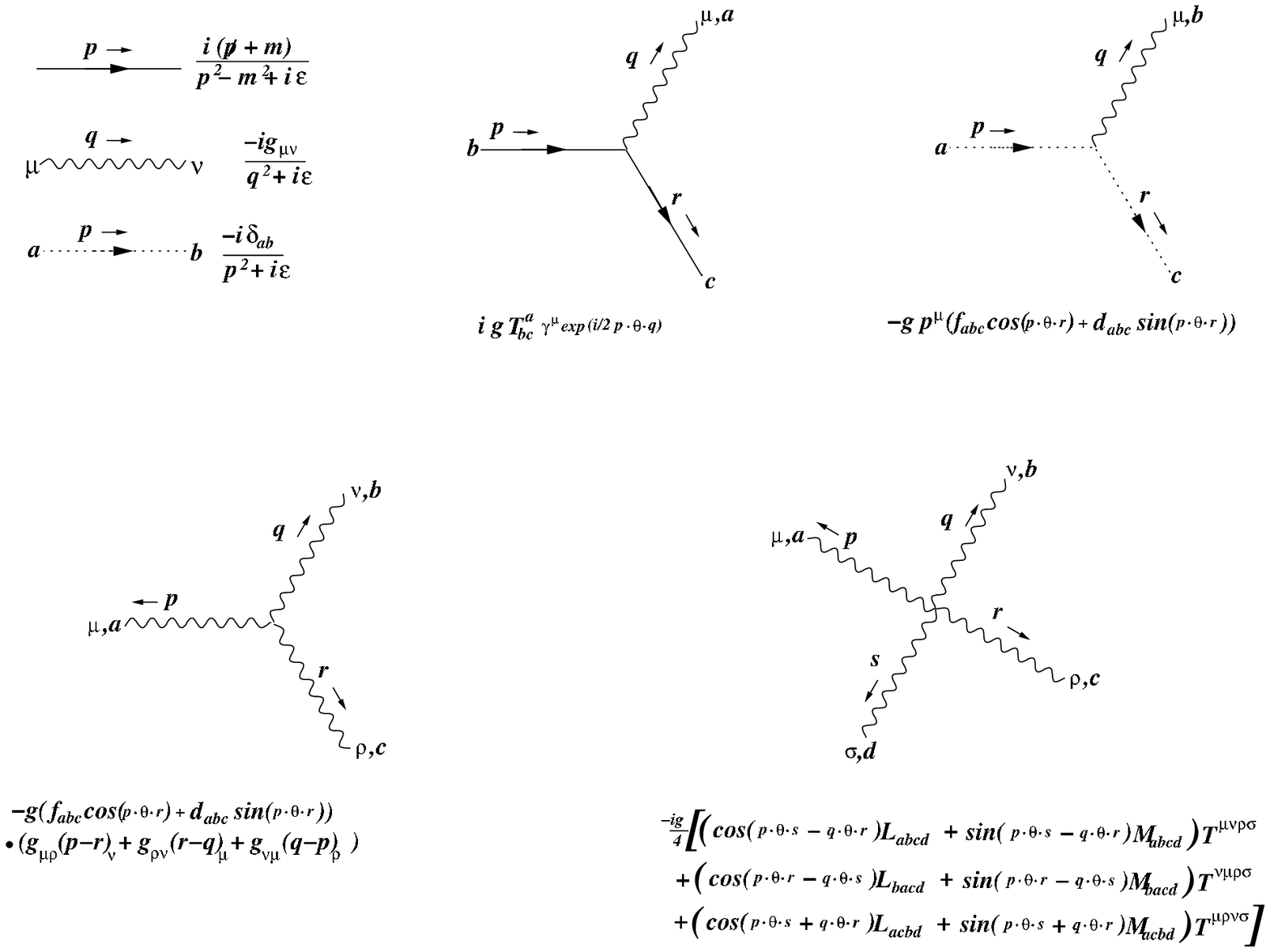}
\caption{Feynman rules for fermions (solid lines), gauge particles (wavy lines),
and ghosts (dotted lines). 
Notation: ~~~
$p,q,r,s$ Momenta ~~~~~  
$\mu,\nu,\rho,\sigma$ Lorentz indices ~~~~~
$a,b,c,d$  gauge indices ~~~~~ $T^a_{bc}$ gauge generator ~~~~~
$f_{abc}$ structure constants for $SU(N)$: $[T_a,T_b] = f_{abc}T^c$ ~~~~~~
$d_{abc}$ structure constants for $SU(N)$: $\{T_a,T_b\} = d_{abc}T^c + {1 \over N}\delta_{ab}$
~~~~~ $L_{abcd} \equiv d_{abe} d_{cde} + d_{ade} d_{cbe} - f_{abe} f_{cde} - f_{ade} f_{cbe} $ ~~~~~
 $M_{abcd} \equiv d_{abe} f_{cde} - d_{ade} f_{cbe} + f_{abe} d_{cde} - f_{ade} d_{bce}$ ~~~~~
$T_{\mu\nu\rho\sigma} \equiv g_{\mu\nu}g_{\rho\sigma} +g_{\mu\sigma}g_{\nu\rho}- 2 g_{\mu\rho}g_{\nu\sigma}$ 
~~~~ For QED/Weak vertices, index $0$ corresponds to a photon: $d_{0,i,j} = \delta_{ij}$, 
 $d_{0,0,i} = 0$, and $d_{0,0,0} = 1$, $f_{0,a,b}$ = 0.}
\label{feynfig}
\end{figure}

\pagebreak

\section{Kaon System}
\label{app:kaon}
The loop function in (\ref{loop}) is given by
\begin{equation}
\begin{array}{l}
f(x) \equiv \frac{x}{(1-x)^2}(1-\frac{11 x}{4} + \frac{x^2}{4} -\frac{3 x^2 ln(x)}{2(1-x)}) \\
f(x,y) \equiv xy \left(\frac{-3}{4(1-x)(1-y)} + \frac{ln(y)(1-2y+\frac{y^2}{4})}{(y-x)(1-y)^2}   
   +  \frac{ln(x)(1-2x+\frac{x^2}{4})}{(x-y)(1-x)^2} \right) \\
\end{array}
\end{equation}
Numerically, 
\begin{equation}
\begin{array}{l}
f((m_t/m_W)^2) \approx 2.5 \\
f((m_c/m_W)^2) \approx 2 \cdot 10^{-4} \\
f((m_c/m_W)^2,(m_t/m_W)^2 ) \approx 2 \cdot 10^{-3}
\end{array}
\end{equation}
In the $\nc$ case with $\eta =0$, the imaginary part of the loop integral for the box graph with
a virtual quark $q$ becomes
\begin{equation}
\begin{array}{l}
\int d^4 k ~  
	\overline{u}(p_1) \gamma_\mu (1-\gamma_5) (\slashchar{p_1}-\slashchar{k}+m_q) 
	\gamma_\nu (1-\gamma_5) d(p_1-k) \\
\times
       \overline{u}(p_2) \gamma_\mu (1-\gamma_5) (\slashchar{k}-\slashchar{p_2}+m_q) 
	\gamma_\nu (1-\gamma_5) d(k-p_2) \\
	\times \frac{(\overline{V}_{qd} \overline{V^*}_{qs})^2}
	 {((p_2+k)^2-m_q^2)((p_1-k)^2-m_q^2) (k^2-{m_W}^2)^2}  \\ 
\end{array}
\end{equation}
which in the high loop momentum limit ($k \gg p_1,p_2$) is approximately 
\begin{equation}
\label{ncloop}
i \lambda^3 \rho \int^{M}_{m_W}{  d^4 k ~ \frac{ (\frac{k^2 m_q^4}{4 m_W^4} + k^2 - \frac{2 m_q^4}{m_W^2})
		\abs{k}\abs{p_1 \cdot \theta \cdot p_2} } {(k^2-m_q^2)^2 (k^2-m_W^2)^2 }}
\end{equation}
where we have introduced the cutoff $M \sim \Lambda$ explicitly since we don't know the theory at higher
energies (taking this limit to infinity doesn't change the answer appreciably.)
The imaginary part of the integral (\ref{ncloop}) for $q=c$ is approximately
\begin{equation}
\Im(nc~loop) \approx \frac{\lambda^2}{m_W^2} \frac{m_K}{18 m_W}
\left( \frac{29 \pi}{2} \xi^2 + \frac{134 \xi^3}{1+\xi^2} - \frac{20 \xi^3}{4 + \xi^2} \right)
\end{equation}
where $\xi \approx \frac{m_W}{\Lambda}$.
For small values of $\xi \ll 1$, this is approximately  
\begin{equation}
\Im(nc~loop) \approx \frac{\lambda^2}{m_W^2} \frac{m_K}{m_W} \xi^2
\end{equation}
which is the simplified form we use in (\ref{ekresult}).
\section{$\epsilon'/\epsilon$}
\label{app:eps}

Direct $\cpviol$ in the SM implies that two or more diagrams contribute
to the kaon decay with disparate weak and strong phases. In
$\ncg$, the vertex phases mimic a weak phase ($\ie$ we use the
ncCKM matrix). To give an estimate for the effects of $\ncg$
on $\epsilon'/\epsilon$, we consider a typical electroweak
penguin loop integral. In the limit of high loop momentum, the penguin
is characterized by the dimensionless number $P_l$
\begin{equation}
P_l \approx \int_{m}^{M}
{\frac{d^4k}{(2\pi)^4}\frac{ ~ i~ m_K ~sin(q\cdot \theta \cdot k)}{k^5}} 
\end{equation}
where $m$ is the mass of the heaviest particle in the loop and $q$ is the typical
momentum of the process $\sim m_t$ in a hadron machine. 
Switching to Euclidean space and performing the integral,
\begin{equation}
P_l \approx \abs{m_K \theta q} \left[Ci(\abs{\theta q \Lambda})
   - Ci(\abs{\theta q m})  \right] + \frac{sin(\abs{\theta q m})}{m} 
- \frac{sin(\abs{\theta q \Lambda})}{\Lambda}
\end{equation}
where we take $M \sim \Lambda$. We use the cosine integral function which for small
values of its argument is
\begin{equation}
Ci(x) \approx const. + ln(x) - \frac{x^2}{4} + \frac{x^4}{4! 4} + \cdot \cdot \cdot
\end{equation}
Taking the average mass $m \sim m_K$ for simplicity, we obtain in the limit
of small $\xi \equiv \theta m_W^2$ 
\begin{equation}
P_l \approx  2\frac{m_K}{m_W} \xi^2~ln(\frac{m_W}{\xi m_K})
\end{equation}
as quoted in (\ref{penguin}).

\end{document}